\documentclass[nofootinbib,prd,twocolumn,showpacs,showkeys,preprintnumbers]{revtex4-1}
\usepackage{hyperref,amssymb,amsmath,mathrsfs,bm,graphicx}
\begin{document}
\title {Tilted shear-free axially symmetric  fluids: silent universes or deaf observers?}
\author{L. Herrera}
\email{lherrera@usal.es}
\affiliation{Instituto Universitario de F\'isica
Fundamental y Matem\'aticas, Universidad de Salamanca, Salamanca 37007, Spain }
\author{A. Di Prisco}
\email{adiprisc@ciens.ucv.ve}
\affiliation{Escuela de F\'\i sica, Facultad de Ciencias, Universidad Central de Venezuela, Caracas 1050, Venezuela}
\author{J.  Carot}
\email{jcarot@uib.cat}
\affiliation{Departament de  F\'{\i}sica, Universitat Illes Balears, E-07122 Palma de Mallorca, Spain}

\date{\today}
\begin{abstract}
We carry on a systematic study of the physical properties of  axially symmetric  fluid distributions, which appear to be  geodesic, shear--free, irrotational, non--dissipative and purely electric, for the comoving congruence of observers,  from the point of view of the tilted congruence. The vanishing of   the magnetic part of the Weyl tensor for  the comoving congruence of observers,  suggests that no gravitational radiation is produced during the evolution of the system. Instead, the magnetic part of the Weyl tensor as measured by tilted observers  is non vanishing (as well as the shear, the four--acceleration, the vorticity and the dissipation), giving rise to a flux of gravitational radiation that can be characterized through the super--Poynting vector. This result strengthens further the relevance of the role of  observers in the description of  a physical system. An explanation of this dual interpretation in terms of the information theory, is provided.
\end{abstract}
\maketitle

\section{Introduction}
In a recent paper \cite{sg} we have analyzed in some detail shear--free and geodesic dissipative fluids, using a general framework for studying axially symmetric dissipative fluids \cite{1}, based on the 1+3 approach \cite{21cil, n1, 22cil, nin}.

Such configurations (which have been previously  considered in great detail by Coley and McManus \cite{c1, c2}), are shown to be  necessarily irrotational and purely electric (the magnetic part of the Weyl tensor vanishes). Therefore, these fluid distributions produce spacetimes which belong to what are known as silent spacetimes \cite{s1, s2, s3}. Strictly speaking the term ``silent universe'' includes additional restrictions, such as the gravitational field is sourced by
dust and cosmological constant only.  However here we shall use this term as implying only the vanishing of the magnetic part of the Weyl tensor and the vorticity.

On the other hand, the magnetic part of the Weyl tensor as well as the vorticity of the fluid lines, are described by tensors defined in terms of the four--velocity of the fluid. Accordingly it is pertinent to ask, if the above mentioned properties (irrotational and purely electric) remain valid for a congruence of observers, tilted (Lorentz boosted) with respect to the  congruence of  comoving observers which, as is obvious, are described by  a different four--velocity vector field.

This issue is related to the well known fact that  there is an observer dependence in the description of the source (see \cite{1t}--\cite{t5} and references therein), related to the arbitrariness in the choice of the four velocity in terms of which the energy--momentum tensor is split, and the kinematical variables are defined.

Thus for example, it can  be shown \cite{38t}, that the usual interpretation of the Lemaitre--Tolman--Bondi spacetime \cite{25, 26, 27}, as geodesic and produced by a non-dissipative dust, is valid for comoving observers exclusively. Tilted observers would detect real (entropy producing)  dissipative processes in such spacetime, and the fluid congruence is no longer geodesic. An explanation for this particular  duality in the interpretation of the physical properties of the fluid, in terms of the information theory, was given in \cite{entropy}.

It is the purpose of this work to analyze in detail  the physical properties of axially symmetric  fluid distributions, which appear to be  geodesic and shear--free, for the comoving congruence of observers,  from the point of view of the tilted congruence. To simplify the analysis we shall consider that the fluid distribution in the comoving frame is non--dissipative. As expected from previous work (see \cite{1t}--\cite{t5} and references therein), the fluid distribution appears to be dissipative for the tilted observer.

The novelty in this work is, as we shall see, that unlike the comoving observers, the tilted ones will detect a flux of gravitational radiation associated to the magnetic part of the Weyl tensor, which for the tilted observers is non vanishing. This is a remarkable result, since the vanishing (or not) of the magnetic part of the Weyl tensor is very often invoked as a significant property  of a given spacetime (see \cite{k2, k1} and references therein). As in \cite{entropy}, an explanation for such a result is given in terms of the information theory. However,  in this work we stress the fact  that an argument similar to the one put forward by Bennet \cite{ben} to solve the Maxwell's demon paradox \cite{max}, may be used to explain the very different pictures of a given system, presented by different congruences of observers  in general relativity.

Also, it is obtained that the fluid for the tilted congruence, appears to be  shearing, non--geodesic and non  irrotational. 

In order to avoid rewriting most of the equations, we shall very often  refer  to \cite{1, sg}.  Thus, we  suggest  that  the reader have at hand these references, when reading this manuscript.

\section{The shear-free, geodesic, axially symmetric  fluid: the comoving picture }
We shall consider  axially and reflection symmetric, non--dissipative  fluid distributions (not necessarily bounded). For such a system the most general line element may be written in ``Weyl spherical coordinates'' as:

\begin{equation}
ds^2=-A^2 dt^2 + B^2 \left(dr^2
+r^2d\theta^2\right)+C^2d\phi^2+2Gd\theta dt, \label{1b}
\end{equation}
where $A, B, C, G$ are positive functions of $t$, $r$ and $\theta$. We number the coordinates $x^0=t, x^1=r, x^2= \theta, x^3=\phi$.

The energy momentum tensor in the ``canonical'' form reads:
\begin{eqnarray}
{T}_{\alpha\beta}&=& (\mu+P) V_\alpha V_\beta+P g _{\alpha \beta} +\Pi_{\alpha \beta},
\label{6bis}
\end{eqnarray}
where as usual, $\mu, P,  \Pi_{\alpha \beta}, V_\beta$ denote the energy density, the isotropic pressure, the anisotropic stress tensor and  the four velocity, respectively.

We emphasize that, so far,  we are considering an Eckart (comoving) frame  where fluid elements are at rest.

If we now impose the shear--free and the geodesic conditions, and assume that the fluid is non--dissipative,  the line element (\ref{1b}) becomes

\begin{equation}
ds^2=-dt^2+B^2(t)\left[dr^2+r^2d\theta ^2+R^2 (r,\theta)d\phi ^2\right].\label{len}
\end{equation}

From regularity conditions at the origin we must require $R(0,\theta)=R^\prime(0,\theta)=R_{,\theta}(0,\theta)=R_{,\theta \theta}(0,\theta)=0$, where prime denotes derivative with respect to $r$. Also it can be shown that   all geodesic and  shear--free fluids, are necessarily irrotational (see \cite{sg} for details). As mentioned in the Introduction, metrics of  this type have been  thoroughly  investigated in \cite{c1, c2}, therefore we shall not enter into a detailed analysis of their properties here.

For our comoving observer  the four--velocity vector reads

\begin{equation}
V^\alpha =\left(1,0,0,0\right); \quad  V_\alpha=\left(-1,0,0,0\right).
\label{m1}
\end{equation}

We shall next define a canonical  orthonormal tetrad (say  $e^{(a)}_\alpha$), by adding to the four--velocity vector $e^{(0)}_\alpha=V_\alpha$, three spacelike unitary vectors (these correspond to the vectors $\bold K, \bold L, \bold S$ in \cite{1})

\begin{equation}
e^{(1)}_\alpha=K_{\alpha}=
(0,B,0,0); \quad  e^{(2)}_\alpha=L_{\alpha}=
\left(0,0,Br,0\right),
\label{7}
\end{equation}

\begin{equation}
 e^{(3)}_\alpha=S_{\alpha}=(0,0,0,BR),
\label{3nb}
\end{equation}
with $a=0,\,1,\,2,\,3$ (latin indices within the round brackets labeling different vectors of the tetrad)

The  dual vector tetrad $e_{(a)}^\alpha$  is easily computed from the condition 
\begin{equation}
 \eta_{(a)(b)}= g_{\alpha\beta} e_{(a)}^\alpha e_{(b)}^\beta, \qquad e^\alpha_{(a)}e_\alpha^{(b)}=\delta^{(b)}_{(a)},
\end{equation}
where $\eta_{(a)(b)}$ denotes the Minkowski metric.

In the above, the tetrad vector $e_{(3)}^\alpha=(1/BR)\delta^\alpha_\phi$ is parallel to
the only admitted Killing vector (it is the unit tangent to the orbits of the
group of 1--dimensional rotations that defines axial symmetry). The other two
basis vectors $e_{(1)}^\alpha,\,e_{(2)}^\alpha$ define the two {\it unique}
directions that are orthogonal to the 4--velocity and to the Killing vector.

For the energy density and the isotropic pressure, we have
\begin{equation}
\mu=T_{\alpha \beta}e^\alpha_{(0)}e^\beta_{(0)},\qquad P=\frac{1}{3}h^{\alpha \beta}T_{\alpha \beta},
\label{eisp}
\end{equation}
where
\begin{equation}
h^\alpha_{\beta}=\delta ^\alpha_{\beta}+V^\alpha V_{\beta},
\label{vel5}
\end{equation}
whereas the anisotropic tensor  may be  expressed through three scalar functions defined as (see \cite{1}, but notice the change of notation):

\begin{eqnarray}
 \Pi _{KL}=e^\alpha_{(2)}e^\beta_{(1)} T_{\alpha \beta} 
, \quad  \label{7P}
\end{eqnarray}

\begin{equation}
\Pi_{I}=\left(2e^{\alpha}_{(1)} e^{\beta}_{(1)} -e^{\alpha}_{(2)} e^{\beta}_{(2)}-e^{\alpha}_{(3)} e^{\beta}_{(3)}\right) T_{\alpha \beta},
\label{2n}
\end{equation}
\begin{equation}
\Pi_{II}=\left(2e^{\alpha}_{(2)} e^{\beta}_{(2)} -e^{\alpha}_{(3)} e^{\beta}_{(3)}-e^{\alpha}_{(1)} e^{\beta}_{(1)}\right) T_{\alpha \beta}.
\label{2nbis}
\end{equation}

In \cite{sg} it was shown, that for the geodesic, shear--free non--dissipative fluid, we have: $\Pi_{KL}=\Pi_{I}=\Pi_{II}=\Pi$, accordingly, the anisotropic tensor may be written in the form:
\begin{equation}
\Pi_{\alpha \beta}=\Pi \left(e^{(1)}_\alpha e^{(1)}_\beta+e^{(2)}_\alpha e^{(2)}_\beta+e^{(2)}_\alpha e^{(1)}_\beta+e^{(1)}_\alpha e^{(2)}_\beta-\frac{2 h_{\alpha \beta}}{3}\right).
\label{2nanis}
\end{equation}

As mentioned before, for the comoving observer, and the line element (\ref{len}), the four--acceleration, the shear and the vorticity vanish, whereas  for the expansion we get:

\begin{equation}
\Theta=\frac{3\dot B}{B},
\label{theta}
\end{equation}
where overdot denotes derivatives with respect to $t$.

\section{The electric and magnetic parts of the Weyl tensor and the super--Poynting vector}
Let us now introduce the electric ($E_{\alpha\beta}$) and magnetic ($H_{\alpha\beta}$) parts of the Weyl tensor ( $C_{\alpha \beta
\gamma\delta}$),  defined as usual by
\begin{eqnarray}
E_{\alpha \beta}&=&C_{\alpha\nu\beta\delta}V^\nu V^\delta,\nonumber\\
H_{\alpha\beta}&=&\frac{1}{2}\eta_{\alpha \nu \epsilon
\rho}C^{\quad \epsilon\rho}_{\beta \delta}V^\nu
V^\delta\,,\label{EH}
\end{eqnarray}
 where $\eta_{\alpha\beta\mu\nu}$ denotes  the Levi-Civita tensor.

In general, for the line element (\ref{1b}), the electric part of the Weyl tensor has only three independent non-vanishing components, whereas only two components define the magnetic part. However, in our case (comoving observers and line element (\ref{len})) the electric part is defined by a single scalar function $\mathcal{E}$, whereas the magnetic part vanishes. Thus  we may  write:

\begin{widetext}
\begin{equation}
E_{\alpha\beta}=\mathcal{E} \left(e^{(1)}_\alpha
e^{(1)}_\beta+ e^{(2)}_\alpha
e^{(2)}_\beta-\frac{2}{3}h_{\alpha \beta}+ e^{(1)}_\alpha
e^{(2)}_\beta+e^{(1)}_\beta
e^{(2)}_\alpha\right), \label{E'}
\end{equation}
\end{widetext}
\noindent

and
\begin{equation}
H_{\alpha\beta}=0.
\end{equation}

Also, from  the Riemann tensor we may define  three tensors $Y_{\alpha\beta}$, $X_{\alpha\beta}$ and
$Z_{\alpha\beta}$ as

\begin{equation}
Y_{\alpha \beta}=R_{\alpha \nu \beta \delta}V^\nu V^\delta,
\label{Y}
\end{equation}
\begin{equation}
X_{\alpha \beta}=\frac{1}{2}\eta_{\alpha\nu}^{\quad \epsilon
\rho}R^\star_{\epsilon \rho \beta \delta}V^\nu V^\delta,\label{X}
\end{equation}
and
\begin{equation}
Z_{\alpha\beta}=\frac{1}{2}\epsilon_{\alpha \epsilon \rho}R^{\quad
\epsilon\rho}_{ \delta \beta} V^\delta,\label{Z}
\end{equation}
 where $R^\star _{\alpha \beta \nu
\delta}=\frac{1}{2}\eta_{\epsilon\rho\nu\delta}R_{\alpha
\beta}^{\quad \epsilon \rho}$  and $\epsilon _{\alpha \beta \rho}=\eta_{\nu
\alpha \beta \rho}V^\nu$.

From the above tensors, we may define  the super--Poynting
vector  by
\begin{equation}
P_\alpha = \epsilon_{\alpha \beta \gamma}\left(Y^\gamma_\delta
Z^{\beta \delta} - X^\gamma_\delta Z^{\delta\beta}\right).
\label{SPdef}
\end{equation}

 In our case,  we may write:
\begin{equation}
 P_\alpha=P_{(1)}e^{(1)}_\alpha+P_{(2)}e^{(2)}_\alpha.\label{SP}
\end{equation}

In the theory of  the super--Poynting vector, a state of gravitational radiation is associated to a  non--vanishing component of the latter (see \cite{11p, 12p, 14p}). This is in agreement with the established link between the super--Poynting vector and the news functions \cite{5p}, in the context of the Bondi--Sachs approach \cite{7, 8}. 

For the comoving observer and the line element (\ref{len}), the magnetic part of the Weyl tensor vanishes identically, implying  at once that $P_{(1)} =P_{(2)}=0$. In other words, no gravitational radiation is detected by the comoving observer .

We shall now proceed to apply a Lorentz boost to our comoving congruence, in order to obtain the tilted one.
\section{The tilted congruence}
In order to obtain the tilted congruence, we have to find the expression for the four--velocity corresponding to this congruence (in the same globally defined coordinate system as in (\ref{len})). For doing that we shall proceed in three steps.

We shall  first perform a (strictly locally defined) coordinate transformation to the  Locally Minkowskian Frame (LMF). 

Denoting  by $\Lambda _\mu ^{\bar\nu}$  the local coordinate transformation matrix, and  by $\bar V^\alpha$ the components of the four--velocity in such LMF, where $\bar x^\alpha$ denotes the Locally Minkowskian coordinates, we have:
\begin{equation}
\bar V^\mu = \Lambda^{\bar\mu}_{\nu} V^\nu,
\label{V}
\end{equation}
where
\begin{equation}
\Lambda^{\bar 0}_0=1;\quad \Lambda^{\bar 1}_1=B;\quad \Lambda^{\bar 2}_2=Br;\quad \Lambda^{\bar 3}_3=BR.
\label{til1}
\end{equation}

Next, let us apply a Lorentz boost to  the LMF associated to  $\bar V^\alpha$, in order to obtain  the (tilted)  LMF with respect to which a fluid element is moving with some non--vanishing three--velocity $\bar v_i$. 

Thus the four--velocity in the tilted LMF is defined by:
\begin{equation}
\tilde {\bar V}_{\beta}=L^{\bar \alpha}_{\bar \beta} \bar V_\alpha,
\label{til2}
\end{equation}
where  $L^{\bar \alpha}_{\bar \beta}$ denotes the corresponding Lorentz matrix.

The boost is  applied along the  two independent directions ($\bar x^1, \bar x^2$), thus we have:
\begin{equation}
L^{\bar 0}_{\bar 0}=\Gamma;\quad L^{\bar 0}_{\bar i}=-\Gamma \bar v_{ i};\quad L^{\bar i}_{\bar j}=\delta {^i}_j+\frac{(\Gamma-1)\bar v_{i} \bar v_{ j}}{\bar v^2},
\label{til3}
\end{equation}
where latin indices $i, j$ run from $1$ to $3$, $\Gamma\equiv \frac{1}{\sqrt{1-\bar v^2}}$, $\bar v^2=\bar v^2_{1}+\bar v^2_{2}$, and $\bar v_{1}$, $\bar v_{2}$ are the two components of the three--velocity of a fluid element as measured by the tilted observer.

Finally, we have to perform a  transformation from the tilted LMF, back  to the (global) frame associated to  the line element (\ref{len}). Such a transformation  is defined by the inverse of  $\Lambda _\mu ^{\bar \nu}$, and produces the four--velocity of the tilted congruence in our globally defined coordinate system, say $\tilde V^\alpha$. This last operation produces:
\begin{equation}
\tilde e^{(0)}_\alpha =\tilde V_\alpha=(-\Gamma, B\Gamma v_1, Br \Gamma v_2, 0);\quad \tilde V^\alpha=(\Gamma, \frac{\Gamma v_1}{B}, \frac{ \Gamma v_2}{Br}, 0).
\label{til4}
\end{equation}

We can also apply the above procedure to obtain the remaining vectors of the tilted tetrad, we find:
\begin{equation}
\tilde e^{(1)}_\alpha=\tilde K_\alpha=\left(-\Gamma v_1, B\left[1+\frac{(\Gamma-1)v^2_1}{v^2}\right],  \frac{Br(\Gamma-1) v_1v_2}{v^2}, 0 \right)\label{vec},
\end{equation}
\begin{equation}
\tilde e^{(2)}_\alpha=\tilde L_\alpha=\left(-\Gamma v_2, \frac{B(\Gamma-1) v_1v_2}{v^2}, Br\left[1+\frac{(\Gamma-1)v^2_2}{v^2}\right] , 0 \right)\label{vecL},
\end{equation}
and 
\begin{equation}
\tilde e^{(3)}_\alpha\equiv e^{(3)}_\alpha=\tilde S_\alpha=(0, 0, 0, BR),\label{vecII}
\end{equation}
where for simplicity we have omitted the bar over the components of the three velocity.

We can now calculate all the kinematical variables for the tilted congruence.

The four acceleration
\begin{equation}
\tilde a_\alpha=\tilde V^\beta \tilde V_{\alpha;\beta},\label{acc}
\end{equation}
may be expressed through two scalar functions as:
\begin{equation}
\tilde a_\alpha=\tilde a_{(1)}  \tilde e^{(1)}_\alpha+\tilde a_{(2)} \tilde e^{(2)}_\alpha.
\label{tacc}
\end{equation}

From (\ref{tacc}) and (\ref{ao})--(\ref{a2}), we can easily find the explicit expressions for the two scalars $\tilde a_{(1)}$ and $\tilde a_{(2)}$.

It is a simple matter to check that if we put $v=0$ ($\Gamma=1$), we obtain $\tilde a_\alpha=0$, as expected.

Next, the shear tensor
\begin{equation}
\tilde \sigma_{\alpha \beta}=\tilde \sigma_{(a)(b)}e^{(a)}_\alpha e^{(b)}_\beta=\tilde V_{(\alpha;\beta)}+\tilde a_{(\alpha}
\tilde V_{\beta)}-\frac{1}{3}\tilde \Theta \tilde h_{\alpha \beta}, \label{acc}
\end{equation}
 may be  defined through two independent tetrad components (scalars)  $\tilde \sigma_{(1)(1)}$ and $\tilde \sigma_{(2)(2)}$, 
defined by:
\begin{equation}
\tilde \sigma_{I}=3\tilde e^\alpha_{(1)} \tilde e^\beta_{(1)} \tilde \sigma_{\alpha \beta},\quad \tilde \sigma_{II}=3\tilde e^\alpha_{(2)} \tilde e^\beta_{(2)} \tilde \sigma_{\alpha \beta}.
\label{sigmat}
\end{equation}

These two scalars may be easily obtained from (\ref{sigmat}) and the expressions for the non--vanishing  coordinate components of the shear tensor displayed in (\ref{s00t})--(\ref{s33t}).

Again, if we go back to the comoving congruence by assuming $v=0$ $(\Gamma=1)$, we get $\tilde \sigma_{\alpha \beta}=0$.

For the vorticity vector defined as:
\begin{equation}
\tilde \omega_\alpha=\frac{1}{2}\,\eta_{\alpha\beta\mu\nu}\,\tilde V^{\beta;\mu}\,\tilde V^\nu=\frac{1}{2}\,\eta_{\alpha\beta\mu\nu}\,\tilde \Omega
^{\beta\mu}\,\tilde V^\nu,\label{vomega}
\end{equation}
where $\tilde \Omega_{\alpha\beta}=\tilde V_{[\alpha;\beta]}+\tilde a_{[\alpha}
\tilde V_{\beta]}$  denotes the vorticity tensor; we find a single component different from zero,  producing:

\begin{equation}
\tilde \Omega_{\alpha\beta}=\tilde \Omega (\tilde e^{(2)}_\alpha \tilde e^{(1)}_\beta -\tilde e^{(2)}_\beta
\tilde e^{(1)}_\alpha),\label{omegaT}
\end{equation}
and
\begin{equation}
\tilde \omega _\alpha =-\tilde \Omega \tilde e^{(3)}_\alpha.
\end{equation}
with the scalar function $\tilde \Omega$ given by
\begin{equation}
\tilde \Omega =-\frac{\Gamma^2}{2}\left(-\frac{v_2^\prime}{B}-\frac{v_2}{Br}-v_1\dot v_2+v_2\dot v_1+\frac{v_{1,\theta}}{Br}\right).
\label{no}
\end{equation}
Obviously in the limit when $v=0$ the vorticity vanishes.

Finally, the expansion scalar, now reads:
\begin{eqnarray}
\tilde \Theta&=&\dot \Gamma+\frac{3\dot B \Gamma}{B}+\frac{(\Gamma v_1)^\prime}{B}+\left(\frac{1}{r}+\frac{R^\prime}{R}\right)\frac{\Gamma v_1}{B}\nonumber \\&+&\frac{\Gamma v_2 R_{,\theta}}{BRr}+\frac{(\Gamma v_2)_{,\theta}}{Br},
\label{tetat}
\end{eqnarray}
which of course reduces to (\ref{theta}) if $v_1=v_2=0$.

In the above equations and hereafter,   primes and dots denote derivatives with respect to $r$ and $t$ respectively.

For the tilted observers the fluid distribution is described by the energy momentum tensor:
\begin{equation}
\tilde T_{\alpha\beta}= (\tilde \mu_+\tilde P) \tilde V_\alpha \tilde V_\beta+\tilde P g _{\alpha \beta} +\tilde \Pi_{\alpha \beta}+\tilde q_\alpha \tilde V_\beta+\tilde q_\beta \tilde V_\alpha.
\label{6bist}
\end{equation}

It should be observed that for the tilted congruence the system may be dissipative, and the anisotropic tensor depends on three scalar functions.
Thus we may write:
\begin{widetext}
\begin{eqnarray}
\tilde \Pi_{\alpha \beta}=\frac{1}{3}(2\tilde \Pi_I+\tilde \Pi_{II})\left(\tilde e^{(1)}_\alpha  \tilde e^{(1)}_\beta -\frac{\tilde h_{\alpha
\beta}}{3}\right)+\frac{1}{3}(2\tilde \Pi _{II}+\tilde \Pi_I)\left(\tilde e^{(2)}_\alpha  \tilde e^{(2)}_\beta -\frac{\tilde h_{\alpha
\beta}}{3}\right)+\tilde \Pi _{KL}\left(\tilde e^{(1)}_\alpha
\tilde e^{(2)}_\beta+\tilde e^{(1)}_\beta
\tilde e^{(2)}_\alpha\right) \label{6bb},
\end{eqnarray}
\end{widetext}
with
\begin{eqnarray}
 \tilde \Pi _{KL}=\tilde e^\alpha_{(1)} \tilde e^\beta_{(2)} \tilde T_{\alpha \beta} 
, \quad \label{7P}
\end{eqnarray}

\begin{equation}
\tilde \Pi_I=\left(2 \tilde e^\alpha_{(1)}  \tilde e^\beta_{(1)}  -  \tilde e^\alpha_{(2)}  \tilde e^\beta_{(2)}- \tilde e^\alpha_{(3)}  \tilde e^\beta_{(3)}\right) \tilde T_{\alpha \beta},
\label{2n}
\end{equation}
\begin{equation}
\tilde \Pi_{II}=\left(2 \tilde e^\alpha_{(2)}  e^\beta_{(2)} -\tilde e^\alpha_{(1)}  \tilde e^\beta_{(1)} -\tilde e^\alpha_{(3)}  \tilde e^\beta_{(3)}\right) \tilde T_{\alpha \beta}.
\label{2nbis}
\end{equation}

Finally,  we may write for the heat flux vector:
\begin{equation}
\tilde q_\mu=\tilde q_{(1)} \tilde  e^{(1)}_\mu+\tilde q_{(2)} \tilde e^{(2)}_\mu.
\label{qn1}
\end{equation}

Since, both congruences of observers are embedded within the same space--time (\ref{len}), then it is obvious that  the Einstein tensor  is the same for both congruences, and therefore the energy--momentum tensors, although split differently, also must be the same.

Then equating  (\ref{6bis}) and (\ref{6bist}), and projecting  on all possible combinations of tetrad vectors (tilted and non--tilted), we  find expressions for the  physical variables measured by comoving observers, in terms of the tilted ones, and viceversa.  These are exhibited in the Appendix B.

For the tilted congruence, the non--vanishing components of  the electric and magnetic parts of the Weyl tensor have been calculated and their expressions are given in the Appendix C. These tensors  may be expressed through the five scalars ($ \tilde{\mathcal{E}}_{I}$, $\tilde{\mathcal{E}}_{II}$, $\tilde{\mathcal{E}}_{KL}$, $\tilde{H}_{1}$, $\tilde{H}_{2}$),  as follows:

\begin{widetext}
\begin{equation}
\tilde E_{\alpha\beta}=\frac{1}{3}\left(2\tilde{\mathcal{E}}_{I}+\tilde{\mathcal{E}}_{II}\right) \left(\tilde e^{(1)}_\alpha
\tilde e^{(1)}_\beta-\frac{1}{3}\tilde h_{\alpha \beta}\right) +\frac{1}{3}\left(2\tilde{\mathcal{E}}_{II}+\tilde{\mathcal{E}}_{I}\right)\left (\tilde e^{(2)}_\alpha
\tilde e^{(2)}_\beta-\frac{1}{3}\tilde h_{\alpha \beta}\right)+\tilde{\mathcal{E}}_{KL} \left(\tilde e^{(1)}_\alpha
\tilde e^{(2)}_\beta+\tilde e^{(1)}_\beta
\tilde e^{(2)}_\alpha\right), \label{E'}
\end{equation}
\end{widetext}
\noindent

and
\begin{equation}
\tilde H_{\alpha\beta}=\tilde{H}_{1}\left(\tilde e^{(1)}_\beta
\tilde e^{(3)}_\alpha+\tilde e^{(1)}_\alpha
\tilde e^{(3)}_\beta \right)+\tilde {H}_{2}\left(\tilde e^{(3)}_\alpha
\tilde e^{(2)}_\beta+\tilde e^{(2)}_\alpha
\tilde e^{(3)}_\beta \right)\label{H'},
\end{equation}
where the above mentioned scalars are expressed through the non--vanishing components  of the electric and magnetic parts of the Weyl tensor, as indicated in the Appendix C.

The above expressions produce for  the super--Poynting vector:
\begin{widetext}
\begin{eqnarray}
\tilde P_{(1)} =
\frac{2\tilde{H}_{2}}{3}\left(2\tilde{\mathcal E}_{II}+\tilde{\mathcal E}_{I}\right)+2\tilde{H}_{1} \tilde{\mathcal E}_{KL}+ 32\pi^2 \tilde q_{(1)}\left(\tilde \mu+\tilde P+\frac{\tilde \Pi_{I}}{3}\right) 
+32\pi^2 \tilde q_{(2)}\tilde \Pi_{KL} ,\nonumber
\\
\tilde P_{(2)}=-\frac{2\tilde H_{1}}{3}\left(2 \tilde {\mathcal E}_{I}+\tilde {\mathcal E}_{II}\right)-2\tilde{H}_{2}\tilde {\mathcal E}_{KL}
+ 32\pi^2 \tilde q_{(2)}\left(\tilde \mu+\tilde P+\frac{\tilde \Pi_{II}}{3}\right)+32\pi^2\tilde q_{(1)}\tilde \Pi_{KL} . \label{SPP}
\end{eqnarray}
\end{widetext}

We can identify two different contributions in (\ref{SPP}). On the one hand we have contributions from the  heat transport process. These are in principle independent of the magnetic part of the Weyl tensor, which explains why they  remain in the spherically symmetric limit.  Next we have contributions related to the gravitational radiation. These contributions are described by the first two terms in $\tilde P_{(1)} $ and $\tilde P_{(2)} $. In order of these contributions to be different from zero we require that, both, the electric and the magnetic part of the Weyl tensor to be non--vanishing. More  specifically,  the sum of the first two terms in $\tilde P_{(1)} $ and $\tilde P_{(2)} $ should not vanish. This is in fact the case, as can be seen from (\ref{esce1})--(\ref{esce3}) and (\ref{h1c}), (\ref{h2c}). Indeed, the vanishing of the above mentioned  terms implies $R\sim r\cos\theta$, which produces the vanishing of the Weyl tensor (conformal flatness). Therefore, excluding the particular conformally flat  case,  the tilted observer detects a non--vanishing gravitational contribution of the super--Poynting vector, which as mentioned above indicates the presence of gravitational radiation.

\section{conclusions}
Using the framework developed in \cite{1} and the results obtained in (\cite{sg}),  we have compared the physical properties of  a physical system described by the line element (\ref{len}), as observed by two different congruences of observers (comoving and tilted). 

Thus, whereas the fluid is shear--free, geodesic, irrotational and non--dissipative, from the point of view of the comoving observer, it appears non--geodesic, shearing, dissipative and endowed with vorticity, for the tilted congruence. 

The fact that tilted observers detect dissipation in a system that appears non--dissipative for comoving observers, is not new and was emphasized in \cite{38t}. To explain such difference in the description of  a given system, as provided by different congruences of observers, it has been conjectured in \cite{entropy} that the origin of this strange situation resides in the fact  that passing from one of the congruences to the other we usually overlook the fact that both congruences of observers store a different amount of information.

This is in fact the   clue to resolve the quandary about the presence or not of dissipative processes, depending
on the congruence of observers, that carry out the analysis of the system.

However, in the present case the difference is still sharper since the tilted observer not only  detect a dissipative process, but also gravitational radiation. Both phenomena are of course absent in the description of the comoving observer. This last point is relevant since the tilted observer also detects vorticity, and as pointed out in \cite{5p}, vorticity  and gravitational radiation are tightly associated.

The explanation for such a difference is basically the same as the one proposed for dissipative processes described by the heat flux vector (remember that gravitational radiation is a dissipative process too), and reminds us the resolution of the well known paradox of the Maxwell's demon \cite{max}. 

The Maxwell's demon (in one of its many, but equivalent presentations) is a small ``being'' living in a cylinder filled with a gas, and divided in two equal portions, by a partition with a small door. Then the demon may  open the door when the molecules come from the right, while closing it when the molecules approach from the left. Doing so the demon is able to concentrate all the molecules on the left, reducing the entropy by $NK\ln2$ (where $N$ is the number of molecules, and $K$ is the Boltzman constant), thereby violating the second law of thermodynamics. Brillouin \cite{bri} tried to solve the paradox by arguing that in the process of selection of molecules, the demon increases the entropy  by an amount equal or larger than the decreasing of entropy achieved by concentrating all molecules on one side. However, soon after, different researchers were able to propose different ways by means of which the demon could select the molecules in a reversible way (i. e. without entropy production).  It has been  necessary to wait for more than a century, until Bennet \cite{ben} gave a satisfactory resolution of this paradox.

Roughly speaking, Bennet showed that the irreversible act which prevents the violation of the second law, is not the selection of molecules to put all of them in one side of the cylinder, but the restauration of the measuring apparatus by means of which the selection is achieved, to the standard state previous to the state where the demon knows from which side comes any molecule. The erasure of such information, according to the Landauer's  principle \cite{lan}, entails dissipation. In other words, to get the demon's mind back to  its initial state, generates dissipation. A somehow similar picture appears when we go from comoving (which assign zero value to the three-velocity of any fluid element) to tilted observers, for whom the three-velocity represents another degree of freedom. The erasure of the information stored by comoving observers (vanishing three velocity), when going to the tilted observers, explains the presence of dissipative processes (included gravitational radiation) observed by the latter. The above comments provide full significance to  the statement by  Max Born: {\it ``Irreversibility is a consequence of the explicit introduction of ignorance into the fundamental laws''} \cite{mb}.

Finally, it is worth mentioning that the effect described here (the detection of gravitational radiation by tilted observers), somehow reminds us the Unruh effect \cite{davies, unruh}, according to which an accelerating observer  (Rindler) in a Minkowski vacuum state will observe a thermal spectrum of particles, thereby indicating that two different sets of observers (inertial and Rindler) describe the same state in very different terms. 

Of course the Unruh effect is of quantum nature, whereas our results belong to the classical realm. However the main morale emerging from both results, points to the same direction, namely: the description of a physical system may  heavily rely on the nature of the observer carrying on the analysis of the system.

\appendix 
\section{Kinematical variables}
The non--vanishing coordinate components of the four--acceleration for the tilted congruence are:
\begin{equation}
\tilde a_0=-\Gamma\left[\dot \Gamma+\frac{\Gamma^\prime v_1}{B}+\frac{\Gamma_{,\theta} v_2}{Br}+\frac{\Gamma v^2 \dot B}{B}\right],
\label{ao}
\end{equation}
\begin{eqnarray}
\tilde a_1&=&\Gamma B\left[ (\Gamma v_1)^{.}+\frac{(\Gamma v_1)^\prime v_1}{B}+\frac{(\Gamma v_1)_{,\theta} v_2}{Br}-\frac{\Gamma v_2^2}{Br}\right.\nonumber \\ &&\left. +\frac{\Gamma v_1 \dot B}{B}\right],
\label{a1}
\end{eqnarray}
\begin{eqnarray}
\tilde a_2=\Gamma Br\left[ (\Gamma v_2)^{.}+\frac{(\Gamma v_2)^\prime v_1}{B}+\frac{(\Gamma v_2)_{,\theta} v_2}{Br}\right.\nonumber \\ \left. +\frac{\Gamma v_2 \dot B}{B}+\frac{\Gamma v_2v_1}{Br}\right].
\label{a2}
\end{eqnarray}

The non--vanishing  coordinate components of the shear tensor are:
\begin{widetext}
\begin{eqnarray}
\tilde \sigma_{00}=-\frac{2\dot \Gamma (1-\Gamma^2)}{3}+\frac{\Gamma^2\Gamma^\prime v_1}{B}+\frac{1-\Gamma^2}{3B}\left[(\Gamma v_1)^\prime+\Gamma v_1\left(\frac{1}{r}+\frac{R^\prime}{R}\right)\right]+\frac{\Gamma^2\Gamma_{,\theta} v_2}{Br}+\frac{1-\Gamma^2}{3Br}\left[(\Gamma v_2)_{,\theta}+\frac{\Gamma v_2R_{,\theta}}{R}\right],
\label{s00t}
\end{eqnarray}
\end{widetext}
\begin{widetext}
\begin{eqnarray}
\tilde \sigma_{01}&=&\frac{B}{2}\left[(\Gamma v_1)^{.} (1-\Gamma^2)-\frac{\Gamma^2 \dot \Gamma v_1}{3}-\frac{\Gamma^\prime(1+\Gamma^2v_1^2)}{B}-\frac{\Gamma^2 v_1(\Gamma v_1)^\prime}{3B} +\frac{\Gamma^3 v_2^2}{Br}+\frac{2\Gamma^3v_1^2}{3B}\left(\frac{1}{r}+\frac{R^\prime}{R}\right)-\frac{\Gamma^2v_1 v_2 \Gamma_{,\theta}}{Br}-\frac{\Gamma^2 v_2( \Gamma v_1)_{,\theta}}{Br}\right.  \nonumber  \\&&\left. +\frac{2\Gamma^2v_1 ( \Gamma v_2)_{,\theta}}{3Br}+
\frac{2\Gamma^3v_1 v_2R_{,\theta}}{3BrR}\right],
\label{s01t}
\end{eqnarray}
\end{widetext}
\begin{widetext}
\begin{eqnarray}
\tilde \sigma_{02}&=&\frac{Br}{2}\left[(\Gamma v_2)^{.} (1-\Gamma^2)-\frac{\Gamma^2 \dot \Gamma v_2}{3}-\frac{\Gamma^\prime \Gamma^2 v_1 v_2}{B}+\frac{2\Gamma^2 v_2(\Gamma v_1)^\prime}{3B} -\frac{\Gamma^3v_1 v_2}{3B}\left(\frac{1}{r}-\frac{2R^\prime}{R}\right)-\frac{\Gamma^2 v_2^2 \Gamma_{,\theta}}{Br}-\frac{\Gamma^2 v_2( \Gamma v_2)_{,\theta}}{3Br}\right.  \nonumber  \\&&\left. -\frac{\Gamma^2v_1 (\Gamma v_2)^{\prime}}{B}-\frac{\Gamma _{,\theta}}{Br}+
\frac{2\Gamma^3 v_2^2R_{,\theta}}{3BrR}\right],
\label{s02t}
\end{eqnarray}
\end{widetext}
\begin{widetext}
\begin{eqnarray}
\tilde \sigma_{11}&=&B^2\left\{-\frac{\dot \Gamma(1+\Gamma^2 v_1^2)}{3}+\Gamma^2 v_1 (\Gamma v_1)^{.}+\frac{(1+\Gamma^2v_1^2)}{3B}\left[2(\Gamma v_1)^\prime  -\Gamma v_1\left(\frac{1}{r}+\frac{R^\prime}{R}\right)\right]-\frac{\Gamma^3 v_2^2 v_1}{Br}\right. \nonumber \\  &+& \left.\frac{\Gamma^2 v_2 v_1( \Gamma v_1)_{,\theta}}{Br}-\frac{(1+\Gamma^2v_1^2)}{3Br}\left[( \Gamma v_2)_{,\theta}+
\frac{\Gamma v_2R_{,\theta}}{R}\right] \right \},
\label{s11t}
\end{eqnarray}
\end{widetext}

\begin{widetext}
\begin{eqnarray}
\tilde \sigma_{12}&=&\frac{B^2r}{2}\left\{-\frac{2\dot \Gamma \Gamma^2 v_1v_2}{3}+\Gamma^2 v_2 (\Gamma v_1)^{.}+ \Gamma^2 v_1 (\Gamma v_2)^{.}+\frac{\Gamma^2v_1 v_2}{3B}\left[(\Gamma v_1)^\prime+\Gamma v_1\left(\frac{1}{r}-\frac{2R^\prime}{R}\right)\right]+\frac{(1+\Gamma^2v_2^2)(\Gamma v_1)_{,\theta}}{Br}\right. \nonumber \\&+ &\left. \frac{1+\Gamma^2 v_1^2}{B}\left[(\Gamma v_2)^\prime-\frac{\Gamma v_2}{r}\right]+\frac{\Gamma^2 v_1 v_2}{3Br}\left[(\Gamma v_2)_{,\theta}-\frac{2\Gamma v_2 R_{,\theta}}{R}\right]\right \},
\label{s12t}
\end{eqnarray}
\end{widetext}

\begin{widetext}
\begin{eqnarray}
\tilde \sigma_{22}&=&B^2r^2\left\{-\frac{\dot \Gamma (1+\Gamma^2 v_2^2)}{3}+\Gamma^2 v_2 (\Gamma v_2)^{.}+\frac{\Gamma^2v_1 v_2}{B}(\Gamma v_2)^\prime+\right. \nonumber \\&+ &\left. \frac{(1+\Gamma^2 v_2^2)}{3B}\left[-(\Gamma v_1)^\prime+\Gamma v_1\left(\frac{2}{r}-\frac{R^\prime}{R}\right)\right]+ \frac{1+\Gamma^2 v_2^2}{3Br}\left[2(\Gamma v_2)_{,\theta}-\frac{\Gamma v_2 R_{,\theta}}{R}\right]\right \},
\label{s22t}
\end{eqnarray}
\end{widetext}

\begin{widetext}
\begin{eqnarray}
\tilde \sigma_{33}=\frac{B^2R^2}{3}\left\{-\dot \Gamma- \frac{1}{B}\left[(\Gamma v_1)^\prime+\Gamma v_1\left(\frac{1}{r}-\frac{2R^\prime}{R}\right)\right]+ \frac{1}{Br}\left[-(\Gamma v_2)_{,\theta}+\frac{2\Gamma v_2 R_{,\theta}}{R}\right]\right \}.
\label{s33t}
\end{eqnarray}
\end{widetext}
\section{Relationships between tilted and non tilted physical variables.}
Proceeding as indicated in section IV,  we get for the tilted variables:
\begin{equation}
\tilde \mu=\Gamma^2\left[\mu+Pv^2+\Pi\left(\frac{v^2}{3}+2v_1v_2\right)\right],
\label{rpv1}
\end{equation}
 
\begin{equation}
\tilde P=P+\frac{\Gamma^2}{3}\left[(\mu+P)v^2+\Pi\left(\frac{v^2}{3}+2v_1v_2\right)\right],
\label{rpv2}
\end{equation}

\begin{widetext}
\begin{eqnarray}
\tilde \Pi_I=\Pi+\Gamma^2\left(\mu+P+\frac{\Pi}{3}\right)(3v_1^2-v^2)+ \frac{2\Pi(\Gamma-1)v_1v_2}{v^2}\left[2-\Gamma+\frac{3(\Gamma-1)v_1^2}{v^2}\right],
\label{rpv3}
\end{eqnarray}
\end{widetext}
\begin{widetext}
\begin{eqnarray}
\tilde \Pi_{II}=\Pi+\Gamma^2\left(\mu+P+\frac{\Pi}{3}\right)(3v_2^2-v^2)+ \frac{2\Pi(\Gamma-1)v_1v_2}{v^2}\left[2-\Gamma+\frac{3(\Gamma-1)v_2^2}{v^2}\right],
\label{rpv4}
\end{eqnarray}
\end{widetext}
\begin{widetext}
\begin{eqnarray}
\tilde \Pi_{KL}=\Gamma^2 v_1v_2(\mu+P)+ \Pi \left[\Gamma +\frac{\Gamma^2v_1v_2}{3}+\frac{2(\Gamma-1)^2 v_1^2 v_2^2}{v^4}\right],
\label{rpv5}
\end{eqnarray}
\end{widetext}

\begin{eqnarray}
-\tilde q_{(1)}&=&\Gamma^2 v_1(\mu+P)+\Pi \Gamma v_1\left[\frac{\Gamma}{3} +\frac{2(\Gamma-1) v_1 v_2}{v^2}\right]\nonumber \\ &+&\Pi\Gamma v_2,
\label{rpv6}
\end{eqnarray}

\begin{eqnarray}
-\tilde q_{(2)}&=&\Gamma^2 v_2(\mu+P)+\Pi \Gamma v_2\left[\frac{\Gamma}{3} +\frac{2(\Gamma-1) v_1 v_2}{v^2}\right]\nonumber \\ &+&\Pi\Gamma v_1.
\label{rpv6}
\end{eqnarray}

Obviously, if in the above we put $v=v_1=v_2=0$, $\Gamma=1$, we obtain at once $\tilde\mu =\mu; \tilde P=P;  \tilde \Pi_I=\tilde \Pi_{II}=\tilde \Pi_{KL}=\Pi$, and $\tilde q_{(1)}=\tilde q_{(2)}=0$, as it must be.

Inversely, we may obtain by the same way, expressions for the physical variables associated to  comoving observers, in terms of the variables corresponding to tilted observers, thus we find:

\begin{equation}
 \mu \Gamma=\tilde \mu\Gamma +\tilde q_{(1)} \Gamma v_1+\tilde q_{(2)}\Gamma v_2,
\label{rpvinv1}
\end{equation}
\begin{widetext} 
\begin{eqnarray}
 3P=\tilde \mu\Gamma^2v^2+\tilde P(3+\Gamma^2v^2)+2\tilde q_{(1)} \Gamma^2 v_1+2\tilde q_{(2)}\Gamma^2 v_2 
+\frac{\tilde \Pi_I \Gamma^2v_1^2}{3}+\frac{\tilde \Pi_{II} \Gamma^2v_2^2}{3}+2\tilde\Pi_{KL}\Gamma^2v_1v_2,
\label{rpvinv2¡inv3}
\end{eqnarray}
\end{widetext}
\begin{widetext}
\begin{eqnarray}
 \Pi=\Gamma^2(\tilde \mu+\tilde P)(v^2-v_1 v_2)+\frac{\tilde \Pi_I}{3}\left\{3+\Gamma^2v_1^2  -\frac{(\Gamma-1)v_1v_2}{v^2}\left[1+\frac{(\Gamma-1)v_1^2}{v^2}\right]\right\}\nonumber \\+\frac{\tilde \Pi_{II}}{3}\left\{3+\Gamma^2v_2^2 -\frac{(\Gamma-1)v_1v_2}{v^2}\left[1+\frac{(\Gamma-1)v_2^2}{v^2}\right]\right\}+\tilde \Pi_{KL}\left[-\Gamma+2\Gamma^2v_1v_2+\frac{2(\Gamma^2-1)v_1^2 v_2^2}{v^4}\right]\nonumber \\+\tilde q_{(1)}\left\{2\Gamma^2v_1-\Gamma v_2\left[1+\frac{2(\Gamma-1)v_1^2}{v^2}\right]\right\}+\tilde q_{(2)}\left\{2\Gamma^2v_2-\Gamma v_1\left[1+\frac{2(\Gamma-1)v_2^2}{v^2}\right]\right\}.
\label{rpv3}
\end{eqnarray}
\end{widetext}
In the limit when $v\rightarrow 0$ the above equations become identities.

\section{The magnetic and the electric parts of the Weyl tensor for the tilted congruence}
Using MAPLE we have calculated the nonvanishing components  of the electric and magnetic part of the Weyl tensor. For the former we found:
\begin{widetext}
\begin{eqnarray}
\tilde E_{00}=\frac{1}{6B^2r^2R(v^2-1)}\left[v_1^2(R^{\prime \prime}r^2-2R^\prime r-2R_{,\theta \theta})+v_2^2(-2R^{\prime \prime}r^2+R^\prime r+R_{,\theta \theta})+6v_1 v_2(R^{\prime}_{,\theta} r-R_{,\theta})\right],
\label{e1}
\end{eqnarray}
\end{widetext}
\begin{widetext}
\begin{eqnarray}
\tilde E_{01}=-\frac{1}{6Br^2R(v^2-1)}\left[v_1(R^{\prime \prime}r^2-2R^\prime r-2R_{,\theta \theta})+3v_2(R^{\prime}_{,\theta} r-R_{,\theta})\right],
\label{e2}
\end{eqnarray}
\end{widetext}
\begin{widetext}
\begin{eqnarray}
\tilde E_{02}=-\frac{1}{6BrR(v^2-1)}\left[v_2(-2R^{\prime \prime}r^2+R^\prime r+R_{,\theta \theta})+3v_1(R^{\prime}_{,\theta} r-R_{,\theta})\right],
\label{e3}
\end{eqnarray}
\end{widetext}
\begin{widetext}
\begin{eqnarray}
\tilde E_{11}=-\frac{1}{6r^2R(v^2-1)}\left[v_2^2(R^{\prime \prime}r^2+R^\prime r+R_{,\theta \theta})+(-R^{\prime \prime}r^2+2R^\prime r+2R_{,\theta \theta})\right],
\label{e4}
\end{eqnarray}
\end{widetext}
\begin{widetext}
\begin{eqnarray}
\tilde E_{12}=\frac{1}{6rR(v^2-1)}\left[v_1v_2(R^{\prime \prime}r^2+R^\prime r+R_{,\theta \theta})+3(R^{\prime}_{,\theta} r-R_{,\theta})\right],
\label{e5}
\end{eqnarray}
\end{widetext}
\begin{widetext}
\begin{eqnarray}
\tilde E_{22}=-\frac{1}{6R(v^2-1)}\left[v_1^2(R^{\prime \prime}r^2+R^\prime r+R_{,\theta \theta})+(2R^{\prime \prime}r^2-R^\prime r-R_{,\theta \theta})\right],
\label{e6}
\end{eqnarray}
\end{widetext}
\begin{widetext}
\begin{eqnarray}
\tilde E_{33}=\frac{R}{6r^2(v^2-1)}\left[v_1^2(2R^{\prime \prime} r^2-R^\prime r-R_{,\theta \theta})+v_2^2(-R^{\prime \prime}r^2+2R^\prime r+2R_{,\theta \theta})\right. \nonumber \\\left.+6v_1v_2(R^\prime_{,\theta}r-R_{,\theta})+ R^{\prime \prime}r^2 +R^\prime r+R_{,\theta \theta}\right].
\label{e7}
\end{eqnarray}
\end{widetext}

These seven components are related by the following four relationships, which allow to write the electric part of the Weyl tensor in terms of three independent scalar functions:

\begin{equation}
Br\tilde E_{00}+r\tilde E_{01}v_1+\tilde E_{02}v_2=0,
\label{coe1}
\end{equation}
\begin{equation}
B^2r^2\tilde E_{00}-\frac{r^2}{R^2}\tilde E_{33}-r^2\tilde E_{11}-\tilde E_{22}=0,
\label{coe2}
\end{equation}
\begin{equation}
\frac{r^2}{R^2}\tilde E_{33}+Br^2\tilde E_{01}v_1+Br\tilde E_{02}v_2+r^2\tilde E_{11}+\tilde E_{22}=0,
\label{coe3}
\end{equation}
\begin{equation}
v_1 v_2(r^2\tilde E_{11}+\tilde E_{22})+Br(r\tilde E_{01}v_2+\tilde E_{02}v_1)+r\tilde E_{12}v^2=0.
\label{coe4}
\end{equation}

Thus we we may express the electric part of the Weyl tensor, in terms of the three scalars ${\mathcal E}_{I}, {\mathcal E}_{II}, {\mathcal E}_{KL}$, given by:
\begin{widetext}
\begin{eqnarray}
\frac{2 \tilde{\mathcal E}_{I}+ \tilde {\mathcal E}_{II}}{3}&=&\Gamma^2v_1^2\tilde E_{00}+\frac{ \tilde E_{11}}{B^2}\left[1+\frac{(\Gamma-1)v_1^2}{v^2}\right]^2+\frac{(\Gamma-1)^2v_1^2 v_2^2 \tilde E_{22}}{B^2r^2v^4}-\frac{\tilde E_{33}}{B^2R^2}+\frac{2\Gamma v_1\tilde E_{01}}{B}\left[1+\frac{(\Gamma-1)v_1^2}{v^2}\right]\nonumber \\&+&\frac{2\Gamma v_1^2v_2(\Gamma-1)\tilde E_{02}}{Brv^2}+\frac{2(\Gamma-1)v_1 v_2\tilde E_{12}}{B^2v^2r}\left[1+\frac{(\Gamma-1)v_1^2}{v^2}\right],
\label{esce1}
\end{eqnarray}
\end{widetext}

\begin{widetext}
\begin{eqnarray}
\frac{2 \tilde{\mathcal E}_{II}+ \tilde {\mathcal E}_{I}}{3}&=&\Gamma^2v_2^2\tilde E_{00}+\frac{ \tilde E_{22}}{B^2r^2}\left[1+\frac{(\Gamma-1)v_2^2}{v^2}\right]^2+\frac{(\Gamma-1)^2v_1^2 v_2^2 \tilde E_{11}}{B^2 v^4}-\frac{\tilde E_{33}}{B^2R^2}+\frac{2\Gamma(\Gamma-1) v_2^2 v_1\tilde E_{01}}{Bv^2}\nonumber \\&+&\frac{2\Gamma v_2\tilde E_{02}}{Br}\left[1+\frac{(\Gamma-1)v_2^2}{v^2}\right]+\frac{2(\Gamma-1)v_1 v_2\tilde E_{12}}{B^2v^2r}\left[1+\frac{(\Gamma-1)v_2^2}{v^2}\right],
\label{esce2}
\end{eqnarray}
\end{widetext}

\begin{widetext}
\begin{eqnarray}
 \tilde {\mathcal E}_{KL}&=&\Gamma^2v_1v_2\tilde E_{00}+\frac{ \tilde E_{22}(\Gamma-1)v_1v_2}{B^2r^2v^2}\left[1+\frac{(\Gamma-1)v_2^2}{v^2}\right]+\frac{(\Gamma-1)v_1 v_2 \tilde E_{11}}{B^2 v^2}\left[1+\frac{(\Gamma-1)v_1^2}{v^2}\right]\nonumber \\&+&\tilde E_{01}\left\{\frac{\Gamma(\Gamma-1) v_2 v_1^2}{Bv^2}+\frac{\Gamma v_2}{B}\left[1+\frac{(\Gamma-1)v_1^2}{v^2}\right]\right\}+\tilde E_{02}\left\{\frac{\Gamma(\Gamma-1) v_2 ^2v_1}{Brv^2}+\frac{\Gamma v_1}{Br}\left[1+\frac{(\Gamma-1)v_2^2}{v^2}\right]\right\}\nonumber \\&+&\tilde E_{12}\left\{\frac{(\Gamma-1)^2 v_2 ^2v_1^2}{B^2rv^4}+\frac{1}{B^2r}\left[1+\frac{(\Gamma-1)v_2^2}{v^2}\right]\left[1+\frac{(\Gamma-1)v_1^2}{v^2}\right]\right\}.
\label{esce3}
\end{eqnarray}
\end{widetext}
Whereas for  the magnetic part  we obtain the following expressions:
\begin{eqnarray}
\tilde H_{03}=\frac{1}{2Br^2(v^2-1)}\left[v_1v_2(R^{\prime \prime}r^2-R^\prime r-R_{,\theta \theta})\right. \nonumber \\\left.+(v_2^2-v^2_1)(R^\prime_{,\theta} r-R_{,\theta})\right],\nonumber \\
\label{h1}
\end{eqnarray}

\begin{equation}
\tilde H_{13}=\frac{1}{2r^2(v^2-1)}\left[v_1(R^{\prime}_{,\theta} r-R_{,\theta})+v_2(R_{,\theta \theta}+R^\prime r)\right],
\label{h2}
\end{equation}

\begin{equation}
\tilde H_{23}=-\frac{1}{2r(v^2-1)}\left[v_1R^{\prime \prime} r^2+v_2(R^\prime_{,\theta} r-R_{,\theta}\right].
\label{h3}
\end{equation}

These  three components  are not independent since they satisfy the relationship:
\begin{equation}
\tilde H_{03}=-\frac{\tilde H_{13} v_1}{B}-\frac{\tilde H_{23}v_2}{Br}.
\label{h4}
\end{equation}

Thus we may express the magnetic part of the Weyl tensor in terms of the two scalars $(\tilde H_1, \tilde H_2)$, given by:
\begin{equation}
\tilde H_1=\frac{\tilde H_{03}(\Gamma-1)v_1}{\Gamma B R v^2}+\frac{\tilde H_{13}}{B^2R},
\label{h1c}
\end{equation}
\begin{equation}
\tilde H_2=\frac{\tilde H_{03}(\Gamma-1)v_2}{\Gamma B R v^2}+\frac{\tilde H_{
23}}{B^2Rr}.
\label{h2c}
\end{equation}

As is obvious from the above expressions, the magnetic part of the Weyl tensor vanishes if we put $v=0$, as it should be since for the comoving congruence the field is purely electric.
\begin{acknowledgments}
This  work  was partially supported by the Spanish  Ministerio de Ciencia e
Innovaci\'on under Research Projects No.  FIS2015-65140-P (MINECO/FEDER).
L.H. thanks      Departament de F\'isica at the  Universitat de les  Illes Balears, for financial support and hospitality. ADP  acknowledges hospitality of the
Departament de F\'isica at the  Universitat de les  Illes Balears. J. C.  gratefully acknowledges financial support from the Spanish Ministerio de Economía y Competitividad through grant ref.: FPA2016-76821.
\end{acknowledgments}


\begin{thebibliography}{100}
\bibitem{sg} L. Herrera, A. Di Prisco,  and  J. Ospino,
{\it Phys.  Rev. D} {\bf  89}, 127502, (2014).
\bibitem{1} L. Herrera, A. Di Prisco, J. Ib\'a\~nez and  J. Ospino,
{\it Phys.  Rev. D} {\bf  89}, 084034, (2014).
\bibitem{21cil} G. F. R. Ellis {\ Relativistic Cosmology} in: Proceedings of the International School of Physics `` Enrico Fermi'', Course 47: General Relativity and Cosmology. Ed. R. K. Sachs (Academic Press, New York and London) (1971).
\bibitem{n1} G. F. R. Ellis and H.  van Ellst, {\it gr--qc/9812046v4} (1998).
\bibitem{22cil} G. F. R. Ellis {\it Gen. Rel. Grav.} {\bf  41}, 581 (2009).
\bibitem{nin} G. F. R. Ellis, R. Maartens and M. A. H. MacCallum, {\it Relativistic Cosmology} (Cambridge U. P., Cambridge) (2012).

\bibitem{c1}  A. A. Coley and D. J. McManus, {\it Class. Quantum Grav.} {\bf 11}, 1261 (1994).
\bibitem{c2}   D. J. McManus and A. A. Coley, {\it Class. Quantum Grav.} {\bf 11}, 2045 (1994).
\bibitem{s1} S. Matarrese, O.  Pantano   and D. Saez, {\it Phys. Rev.} D {\bf
47}  1311 (1993).

\bibitem{s2} M. Bruni, S.  Matarrese  and O. Pantano, 
{\it Astrophys. J.} {\bf 445} 958 (1995). 
\bibitem{s3} H. van Ellst, C. Uggla, W. M. Lesame, G. F. R. Ellisand and R. Maartens,  {\it Class Quantum Grav} {\bf 14} 1151 (1997). 
\bibitem{1t} A. R. King and G. F. R. Ellis {\it Commun. Math. Phys.} {\bf 31}, 209 (1973).
\bibitem{2t} B. O. J. Tupper,  {\it  J. Math. Phys.}  {\bf 22}, 2666 (1981).
\bibitem{4t} A. A. Coley and B. O. J. Tupper {\it Astrophys. J} {\bf 271}, 1 (1983).
\bibitem{5t}  A. A. Coley and B. O. J. Tupper {\it Gen. Rel. Grav.} {\bf 15}, 977  (1983).
\bibitem{6t} B. O. J. Tupper,  {\it Gen. Rel. Grav.} {\bf 15}, 849 (1983).
\bibitem{7t}  A. A. Coley and B. O. J. Tupper {\it Phy. Lett. A} {\bf 100}, 495  (1984).
\bibitem{9t} A. A. Coley, Astrophys. J. {\bf 318}, 487 (1987).
\bibitem{38t} L. Herrera, A. Di Prisco and  J.  Ib\'a\~nez, {\it Phys. Rev. D}  {\bf 84}, 064036 (2011).

\bibitem{2St} L. Herrera, A. Di Prisco,  J.  Ib\'a\~nez and J. Carot, {\it Phys. Rev. D}  {\bf 86}, 044003 (2012).
\bibitem{t1}C. G. Tsagas and M. I. Kadiltzoglou, {\it Phys. Rev. D} {\bf 88}, 083501 (2013).
\bibitem{t2}R. Kumar and S. Srivastava, {\it Astrophys. Space Sci.} {\bf 346}, 567 (2013).
\bibitem{t3}M. Sharif and H. Tahir, {\it Astrophys. Space Sci.} {\bf 351}, 619 (2014).
\bibitem{t3B}J. Fern\'andez and J. Pascual--S\'anchez, {\it Procc. Math. Stat.} {\bf 60}, 361, (2014).
\bibitem{t4}M. Sharif and M. Zaeem Ul Haq Bhatti, {\it Int. J. Mod. Phys. D} {\bf 24}, 1550014 (2015).
\bibitem{t5} Z.Yousaf, K. Bamba and M. Zaeem Ul Haq Bhatti, {\it Phys. Rev. D} {\bf 95}, 024024 (2017).

\bibitem{25} G. Lema\^{\i}tre {\it Ann. Soc. Sci. Bruxelles} {\bf A 53}, 51 (1933) ({\it Gen. Rel. Grav.} {\bf 29}, 641 (1997)) .

\bibitem{26} R. C. Tolman {\it Proc. Natl. Acad Sci} {\bf 20}, 169  (1934) ({\it Gen. Rel. Grav.} {\bf 29}, 935 (1997)).

\bibitem{27} H. Bondi {\it  Mon. Not. R. Astron. Soc.} {\bf 107}, 410 (1947) ({\it Gen. Rel. Grav.} {\bf 31}, 1783 (1999)).
\bibitem{entropy}  L. Herrera, {\it  Entropy} {\bf  19}, 110, (2017).
\bibitem{k2}R. Kumar and S. K. Srivastava, {\it Int. J. Geom. Methods Mod. Phys.} {\bf 11}, 1450043 (2014).
\bibitem{k1}R. Kumar, S. K. Srivastava and V. C, Srivastava, {\it Int. J. Geom. Methods Mod. Phys.} {\bf 12}, 1550103 (2015).
\bibitem{ben} C. H. Bennet, {\it Int. J. Theor. Phys.} {\bf 21}, 905 (1982).
\bibitem{max} J. C. Maxwell, {\it Theory of heat}, (D. Appleton and Co., New York) (1872).
\bibitem{11p} L. Bel, {\it C. R. Acad. Sci.} {247}, 1094 (1958).
\bibitem{12p} L. Bel, {\it Cah.de Phys.} {\bf 16} 59 (1962);  {\it Gen. Rel.Grav.} {\bf 32}, 2047 (2000).
\bibitem{14p}  A. Garc\'ia--Parrado G\'omez Lobo, {\it Class. Quantum Grav.} {\bf 25}, 015006 (2008).
\bibitem{5p} L. Herrera, W. Barreto, J. Carot and A. Di Prisco, {\it Class. Quantum.Grav.} {\bf 24}, 2645 (2007).
\bibitem{7}H. Bondi, M. G. J. van der Burg and A. W. K. Metzner, {\it  Proc. Roy.Soc. A} {\bf 269}, 21 (1962).
\bibitem{8}R. Sachs, {\it Proc. Roy.Soc. A} {\bf 270}, 103 (1962).
\bibitem{bri} L. Brillouin, {\it Science and information theory}, (Academic Press, London) (1956).
\bibitem{lan} R. Landauer, {\it IBM J. Res. Develop.} {\bf 5}, 183 (1961).
\bibitem{mb} M. Born, {\it Natural Philosophy of Cause and Chance}, (Clarendon Press, Oxford) (1949).
\bibitem{davies} P. C. W. Davies, {\it J. Phys. A: Mathematical and General}  {\bf  8}, 609 (1975).
\bibitem{unruh} W. G. Unruh, {\it Phys. Rev. D} {\bf 14}, 870 (1976).
\end{thebibliography}
\end{document}